\documentclass[aps,prb,reprint,superscriptaddress]{revtex4-1}

\usepackage{bm}
\usepackage{amsmath}
\renewcommand{\vec}{\bm}

\begin{document}

%\preprint{}

\title{Can electrostatic field lift spin degeneracy?}


\author{T. G. Tenev}
\email{tenev@phys.uni-sofia.bg}
%\thanks{}
\affiliation{Department of Physics, Sofia University, 5 James Bourchier Blvd, Sofia 1164, Bulgaria}
\author{N. V. Vitanov}
\affiliation{Department of Physics, Sofia University, 5 James Bourchier Blvd, Sofia 1164, Bulgaria}


\date{\today}

\begin{abstract}
There are two well known mechanisms which lead to lifting of energy spin degeneracy of single electron systems - magnetic field and spin-orbit coupling. We investigate
the possibility for existence of a third mechanism in which electrostatic field can lead to lifting of spin-degeneracy directly without the mediation of
spin-orbit coupling. A novel argument is provided for the need of spin-orbit coupling different from the usual relativistic considerations.
It is shown that due to preserved translational invariance spin splitting purely by electrostatic field is not possible for Bloch states. A possible
lifting of spin degeneracy by electrostatic field characterized by broken inversion and broken translational invariance is considered.
\end{abstract}


\maketitle

\section{Introduction}
The field of spintronics\cite{Zutic}, a term coined by Wolf\cite{Wolf}, has attracted considerable attention in the past 20 years. This interest was initially triggered by the
discovery of the effect of giant magnetoresistance (GMR)\cite{Fert,Grunberg} which now has established commercial applications. A parallel interest in semiconductor spintronics has
arisen due to the proposal for a ballistic spin-transistor\cite{DattaDas} which has recently been realized\cite{SFETRealization}. A generalization of the device operating in diffusive
regime has been proposed\cite{Schliemann} as well as other devices \cite{Zutic}. They depend on the spin splitting caused by the spin-orbit coupling through several different mechanisms
usually classified in two big groups. Historically the first group often referred to as Dresselhaus spin-orbit coupling\cite{DresselhausBand,Kane,Weiler78}, or alternatively,  bulk inversion
asymmetry\cite{Zawadski,Winkler,Fabian} (BIA), has as a necessary condition the broken space inversion invariance of a bulk crystal. It also appears in a modified
form\cite{Zutic,Zawadski,Winkler,Eppenga88,Silva92} in quasi two-dimensional (Q2D) structures grown in crystals lacking inversion symmetry. The second group referred to as either Rashba
spin-orbit coupling\cite{Rashba,Ohkawa74,Lassnig} or structure inversion asymmetry\cite{Zutic,Zawadski,Winkler} (SIA) leads to spin splitting in systems lacking macroscopic inversion
symmetry. In all these cases the inversion asymmetry is merely a necessary condition and not a sufficient one. The other necessary condition is taking spin-orbit coupling into account
in various ways but usually through the folding down procedure\cite{Zawadski,Winkler,Fabian,Lassnig} from 8 band or 14 band models. Here we consider whether the breaking of inversion
invariance of electrostatic field in several classes of systems can also be a sufficient condition for lifting of spin degeneracy.

We examine a symmetry argument\cite{Zawadski,Winkler,Fabian,Kittel} due to Kittel\cite{Kittel}. We provide novel derivation of the symmetry argument\cite{Zawadski,Winkler,Fabian,Kittel} relying only on the commutation properties of the time-reversal $\hat{K}$ and space inversion
$\hat{I}$ operators with the translation $\hat{T}_{\vec{R}_n}$ and spin operators $\hat{S}_{\vec{u}}$. This is unlike the usual derivation\cite{Kittel} relying on explicit form of Bloch states. We show a novel argument for the need for introduction of a
spin-orbit coupling term similar to the argument motivating Maxwell to introduce the displacement current term in the Maxwell's equations. We note that the symmetry
argument \cite{Zawadski,Winkler,Fabian,Kittel} involving time-reversal and space inversion invariance does not require the presence of a spin-orbit coupling term in the Hamiltonian but merely to take the spin degree of freedom of the electron into account. The application
of this symmetry argument, to a nonrelativistic model of electron with spin moving in electrostatic field characterized by discrete translational invariance, naturally
suggests the hypothesis that electrostatic field with broken space inversion symmetry can lead to lifting of spin degeneracy. We explore the hypothesis by
perturbation method treatment and show that if discrete translational invariance is preserved electrostatic field can not lift spin-degeneracy. If both translational
and space inversion symmetry of a perturbing electrostatic field are broken a possible contribution to spin splitting by electrostatic field is indicated as far as the perturbation method is applicable. Physically this would be naturally explained since we know\cite{Messiah}%,Hamermesh}
 that every non-accidental degeneracy stems from symmetries of the underlying system and in general reduction of symmetries leads to lifting of degeneracies.

In Sec.~\ref{SEC:TheoreticModel} we introduce the theoretical models. The basic symmetry argument is presented in Sec.~\ref{SUBSEC:SymmetryArgument}.
The transformation properties of Bloch states in a model neglecting spin-orbit coupling but taking
into account the spin degree of freedom are given in
Sec.~\ref{APP:TranBlochStates}. They are used in Sec.~\ref{APP:Spin Degeneracy} to show the appearance of at least four-fold degeneracy in the spectrum
as a consequence of the combination of time-reversal and space inversion invariance and in Sec.~\ref{APP:BrokenSpinDegeneracy} to show the lifting of spin degeneracy
of Bloch states as a consequence of broken space inversion invariance. How the presented symmetry argument differs from the usual one\cite{Zawadski,Winkler,Fabian,Kittel} is discussed in Sec.~\ref{SUBSEC:SymmetryDiscussion}. In Sec.~\ref{SEC:NewArgument} a new argument for the introduction of spin-orbit coupling is presented different from the usual relativistic arguments\cite{Winkler,Messiah,AdvancedSakurai,Thaller}. We then explore the hypothesis formulated in Sec.~\ref{SUBSEC:SymmetryDiscussion} by perturbation method treatment in Sec.~\ref{SEC:PerturbationMethod}. In Sec.~\ref{SEC:Discussion} we discuss certain aspects of the utilized model and methods and we present our conclusions in Sec.~\ref{SEC:Conclusions}.
%% FOR THE COVER LETTER: write down that as far as you are aware this
% derivation deserves to be put in the manuscript for two reasons
%  1) to provide better self-containment of the description and to allow for a reader novel to the field not to look for too many sources for the proofs
%  2) because as far as you are aware exactly this method of proof is not present in any text. Kittel and others use a method in which the explicit form of Bloch functions is used.
%     On the other hand here we use just Bloch theorem.
%  3) Furthermore this method of derivation emphasizes the role of translational invariance.

% If necessary the appendices will be removed from the manuscript if the referees decides that they add nothing new to the discussion and thus just must be removed. However in all
% cases the appendices will remain part of the chapter of the dissertation since a dissertation needs to be more self-contained than a paper and repetition of other more or less
% well known results is more allowed in a dissertation compared with a paper. And I am trying to write down the two things at the same time.

\section{Theoretical Models}\label{SEC:TheoreticModel}
We first consider two basic models of a nonrelativistic electron moving in a pure electrostatic field characterized by discrete translational invariance. We focus our attention on 3D models of the triclinic crystal system. The two classes of models differ from each other by the properties of the electrostatic potential with respect to space inversion symmetry. In both models we take spin degree of freedom into account but neglect the spin-orbit coupling term. The eigenstates of both models are Bloch states\cite{Kittel}, which can be characterized by two quantum numbers: (i) crystal wavevector and (ii) spin index which is the eigenvalue of the $\vec{u}$-component $\hat{S}_{\vec{u}}$ of the spin vector operator $\vec{\hat{S}}$. Furthermore we employ the standard Born-von Karman periodic boundary conditions.
% $\hat{S}_{\vec{u}}|\vec{k},s_{\vec{u}}\rangle=s_{\vec{u}}|\vec{k},s_{\vec{u}}\rangle$.

The first class of models represents 3D crystals of the triclinic pinacoidal symmetry class. Its Hamiltonian takes the form
\begin{equation}
\hat{H}_0=\frac{\hat{\vec{p}}^2}{2m}\hat{\sigma}_0+V_0(\vec{r})\hat{\sigma}_0\; .\label{EQ:BasicModel}
\end{equation}
This represents a nonrelativistic electron moving in a pure electrostatic potential $\phi_0(\vec{r})$ with potential energy
$V_0=-e\phi_0(\vec{r})$ characterized by translational invariance, $[\hat{T}_{\vec{R}_n},\phi_0(\vec{r})]=0$, time-reversal invariance $[\hat{K},\phi_0(\vec{r})]=0$, and space inversion invariance $[\hat{I},\phi_0(\vec{r})]=0$.

The second class of models represents 3D crystals of the triclinic pedial symmetry class in which the overall potential does not possess space inversion invariance. A general potential not possessing space inversion invariance can be split into parts symmetric with respect to space inversion, that is one that commutes with the space inversion operator $\hat{I}$, and a part antisymmetric with respect to space inversion, that is one that anticommutes with $\hat{I}$. We consider the electrostatic field of the second class of models as made of the symmetric part $\phi_{0}(\vec{r})$ and an antisymmetric one $\phi(r)$, satisfying $\hat{I}\phi(r)\hat{I}^+=-\phi(r)$. The Hamiltonian takes the form
\begin{equation}
\hat{H}=\frac{\hat{\vec{p}}^2}{2m}\hat{\sigma}_0+V_0\hat{\sigma}_0 - e\phi(\vec{r})\hat{\sigma}_0\; ,\label{EQ:HamBrokenInversion}
\end{equation}
 While the Hamiltonian in such systems possesses discrete translational invariance
$[\hat{H},\hat{T}_{\vec{R}_n}]=0$, and time-reversal invariance, $[\hat{H},\hat{K}]=0$, it is no longer invariant with respect to space inversion, $[\hat{H},\hat{I}]\neq0$.

In order to emphasize that we have taken spin degree of freedom into account we have written the Hamiltonians (\ref{EQ:BasicModel}) and (\ref{EQ:HamBrokenInversion}) with the $2\times2$ identity matrix $\sigma_0$.

\section{Symmetry Argument}\label{SUBSEC:SymmetryArgument}
\subsection{Transformation of Bloch States}\label{APP:TranBlochStates}
We consider a general Bloch state $|\vec{k},s_{\vec{u}}\rangle$ the properties of which are identical in the two models considered. It satisfies the Bloch theorem\cite{Kittel},
\begin{equation}
\hat{T}_{\vec{R}_n}|\vec{k},s_{\vec{u}}\rangle=e^{-i\vec{k}\cdot\vec{R}_n}|\vec{k},s_{\vec{u}}\rangle\; ,\label{EQ:Bloch}
\end{equation}
where we use the active convention\cite{Tung,Messiah} for the space translation operator $\hat{T}_{\vec{R}_n}$.
The time-reversal transformed state $|\vec{k}',s_{\vec{u}}'\rangle=\hat{K}|\vec{k},s_{\vec{u}}\rangle$ of a Bloch state $|\vec{k},s_{\vec{u}}\rangle$ is still a Bloch state
because the time-reversal operator commutes\cite{Messiah} with
the spatial translation operators $\hat{T}_{\vec{R}_n}$. Applying the time-reversal operator $\hat{K}$ to the Bloch theorem (\ref{EQ:Bloch}), taking
into account that $[\hat{T}_{\vec{R}_n},\hat{K}]=0$ and that $\hat{K}$ as antilinear operator does not commute with complex scalars
$c$ but satisfies the identity\cite{Messiah} $\hat{K}c=c^{*}\hat{K}$, one
obtains the identity $\hat{T}_{\vec{R}_n}|\vec{k}',s_{\vec{u}}'\rangle=e^{i \vec{k}\cdot\vec{R}_n}|\vec{k}',s_{\vec{u}}'\rangle$. Comparing it with the Bloch theorem,
$\hat{T}_{\vec{R}_n}|\vec{k}',s_{\vec{u}}'\rangle=e^{-i\vec{k}'\cdot\vec{R}_n}|\vec{k}',s_{\vec{u}}'\rangle$,
shows that $k'=-k$. Applying $\hat{K}$ to the relation $\hat{S}_{\vec{u}}|\vec{k},s_{\vec{u}}\rangle=s_{\vec{u}}|\vec{k},s_{\vec{u}}\rangle$ and using the relation
$\hat{K}\hat{S}_{\vec{u}}=-\hat{S}_{\vec{u}}\hat{K}$, which follows directly from the definition\cite{Messiah} of the time-reversal
operator $\hat{K}$, one obtains $\hat{S}_{\vec{u}}\hat{K}|\vec{k},s_{\vec{u}}\rangle=-s_u\hat{K}|\vec{k},s_{\vec{u}}\rangle$. Thus the $\hat{K}$-transformed state
$|\vec{k}',s_{\vec{u}}'\rangle=\hat{K}|\vec{k},s_{\vec{u}}\rangle$ is an eigenstate of $\hat{S}_u$ with an eigenvalue $-s_u$, therefore $s_u'=-s_u$.
Summarizing, the time-reversal operator $\hat{K}$ transforms a Bloch state $|\vec{k},s_{\vec{u}}\rangle$ representing an electron moving
with a wavevector $\vec{k}$ and a spin pointing "up" the axis $\vec{u}$ into the Bloch state $|-\vec{k},-s_{\vec{u}}\rangle$,
\begin{equation}
\hat{K}|\vec{k},s_{\vec{u}}\rangle=|-\vec{k},-s_{\vec{u}}\rangle\; ,\label{EQ:KPsi}
\end{equation}
representing an electron moving in the opposite direction with a crystal wavevector $-\vec{k}$ and a spin pointing "down" the axis
$\vec{u}$.

The space translation and space inversion operators do not commute but satisfy\cite{Tung} the identity
\begin{equation}
\hat{I}\hat{T}_{\vec{R}_n}=\hat{T}_{-\vec{R}_n}\hat{I}. \label{EQ:TI}
\end{equation}
Applying the space inversion operator $\hat{I}$ to the Bloch theorem, (\ref{EQ:Bloch}),
using Eq.(\ref{EQ:TI}) and the fact that $\hat{I}$ does not act on the phase factor $e^{-i \vec{k}\cdot\vec{R}_n}$
one obtains
\begin{equation}
\hat{T}_{-\vec{R}_n}\hat{I}|\vec{k},s_{\vec{u}}\rangle=e^{-ikR_n}\hat{I}|\vec{k},s_{\vec{u}}\rangle\; .   \label{EQ:TIPsi}
\end{equation}
Since the $\hat{I}$-transformed Bloch state $I|\vec{k},s_{\vec{u}}\rangle$ satisfies the Bloch theorem it is still a Bloch state,
but in general with different quantum numbers $|\vec{k}',s_{\vec{u}}'\rangle=\hat{I}|\vec{k},s_{\vec{u}}\rangle$. Comparing Eq.(\ref{EQ:TIPsi}) with
the Bloch theorem $\hat{T}_{-\vec{R}_n}|\vec{k}',s_{\vec{u}}'\rangle=e^{ ik' R_n}|\vec{k}',s_{\vec{u}}'\rangle$ one obtains that $k'=-k$. By definition\cite{Messiah}
the space inversion operator $\hat{I}$ commutes with any component of the spin vector operator $\hat{\vec{s}}$. As a consequence,
using the usual procedure applied above to $\hat{S}_{\vec{u}}|\vec{k},s_{\vec{u}}\rangle=s_{\vec{u}}|\vec{k},s_{\vec{u}}\rangle$ gives that $s_{\vec{u}}'=s_{\vec{u}}$. Therefore, the
space inversion operator $\hat{I}$ maps a Bloch state $|\vec{k},s_{\vec{u}}\rangle$, describing electron motion with crystal wavevector
$\vec{k}$ and spin pointing in the direction of axis $\vec{u}$, into the state
\begin{equation}
\hat{I}|\vec{k},s_{\vec{u}}\rangle=|-\vec{k},s_{\vec{u}}\rangle\; ,\label{IPsi}
\end{equation}
representing an electron motion with the same orientation of spin but moving in the opposite direction with a wavevector $-\vec{k}$.
Using the definition of the conjugation operator\cite{Kittel} $\hat{C}=\hat{K}\hat{I}$ and Eqs.~(\ref{EQ:KPsi}) and (\ref{IPsi}),
one obtains the action of the $\hat{C}$ on Bloch states,
\begin{equation}
\hat{C}|\vec{k},s_{\vec{u}}\rangle=|\vec{k},-s_{\vec{u}}\rangle\; .\label{EQ:CPsi}
\end{equation}

\subsection{Spin Degeneracy}\label{APP:Spin Degeneracy}
Supposing the spectrum problem of the Hamiltonian $\hat{H}_0$ with space-inversion invariant electrostatic potential
$\phi_{0}(r)$ solved its eigenvalue-eigenvector problem takes the form of the identity
\begin{equation}
\hat{H}_0|\vec{k},s_{\vec{u}}\rangle\equiv E^0_{\vec{k},s_{\vec{u}}}|\vec{k},s_{\vec{u}}\rangle\; .\label{EQ:H0Eigen}
\end{equation}
The eigenvalues $E^0_{\vec{k},s_{\vec{u}}}$ of $\hat{H}_0$ are labeled with quantum numbers $\vec{k},s_{\vec{u}}$ and the eigenstates of $H_0$ and
$|\vec{k},s_{\vec{u}}\rangle$ satisfying the Bloch theorem possess all the properties of Bloch states, in particular
Eq.~(\ref{EQ:KPsi}) and Eq.~(\ref{EQ:TIPsi}).

Applying the time-reversal operator $\hat{K}$ to Eq.(\ref{EQ:H0Eigen}) and using the time-reversal invariance of the
Hamiltonian $[\hat{H},\hat{K}]=0$, we obtain
\begin{equation}
\hat{H}_0\hat{K}|\vec{k},s_{\vec{u}}\rangle\equiv E^0_{k,s_{\vec{u}}}\hat{K}|\vec{k},s_{\vec{u}}\rangle\; . \label{EQ:HOKTransformed}
\end{equation}
Due to the time-reversal invariance of $\hat{H}_0$ the two linearly independent states $|\vec{k},s_{\vec{u}}\rangle$ and
$|-\vec{k},-s_{\vec{u}}\rangle=\hat{K}|\vec{k},s_{\vec{u}}\rangle$ correspond to the same eigenvalue of $\hat{H}_0$, which we now denote as
$E^0_{\theta}\equiv E^0_{\vec{k},s_{\vec{u}}}=E^0_{-\vec{k},-s_{\vec{u}}}$. The pairs of linearly independent states $(|\vec{k},s_{\vec{u}}\rangle,|-\vec{k},-s_{\vec{u}}\rangle)$
and $(|-\vec{k},s_{\vec{u}}\rangle,|\vec{k},-s_{\vec{u}}\rangle)$ span respectively the 2D subspaces $\varepsilon_{\theta}^0$ and $\varepsilon_{-\theta}^0$
of the Hilbert space of the single-particle system. The 2-fold degeneracy of the eigenenergies $E^0_{\theta}$ and $E^0_{-\theta}\equiv E^0_{-\vec{k},s_{\vec{u}}} = E^0_{\vec{k},-s_{\vec{u}}}$, to which the subspaces $\varepsilon_{\theta}^0$ and $\varepsilon_{-\theta}^0$ correspond, is a consequence
of the time-reversal invariance of $\hat{H}_0$ and is the realization of Kramers degeneracy\cite{Messiah,Sakurai} in the
system described by Eq.~(\ref{EQ:H0Eigen}). By their construction the subspaces $\varepsilon_{\theta}^0$ and $\varepsilon_{-\theta}^0$ are invariant with respect to the time reversal symbolically written as $\hat{K}\varepsilon^0_{\pm\theta}=\varepsilon^0_{\pm\theta}$.
Comparing the spectrum equation $\hat{H_0}|-\vec{k},-s_{\vec{u}}\rangle=E^0_{-\vec{k},-s_{\vec{u}}}|-\vec{k},-s_{\vec{u}}\rangle$ for a state $|-\vec{k},-s_{\vec{u}}\rangle$
with Eq.~(\ref{EQ:HOKTransformed}) allows us to express the Kramers degeneracy in the studied system in the form
\begin{equation}
E^0_{\vec{k},s_{\vec{u}}}=E^0_{-\vec{k},-s_{\vec{u}}} \; .\label{EQ:KDegeneracy}
\end{equation}

Acting on the left of Eq.~(\ref{EQ:H0Eigen}) with the space-inversion operator $\hat{I}$,using the hypothesis
$[\hat{H},\hat{I}]=0$  and the result $|-\vec{k},s_{\vec{u}}\rangle\equiv\hat{I}|\vec{k},s_{\vec{u}}\rangle$ from Eq.~(\ref{IPsi}), one obtains the
identity
\begin{equation}
\hat{H}_0|-\vec{k},s_{\vec{u}}\rangle\equiv E^0_{\vec{k},s_{\vec{u}}}|-\vec{k},s_{\vec{u}}\rangle\; .\label{EQ:H0ITransformed}
\end{equation}
It shows that the space-inversion invariance of the Hamiltonian $\hat{H}_0$ requires that the Bloch states
$|\vec{k},s_{\vec{u}}\rangle$ and $|-\vec{k},s_{\vec{u}}\rangle$ belong to the same eigenenergy,
\begin{equation}
E^0_{k}\equiv E^0_{\vec{k},s_{\vec{u}}}=E^0_{-\vec{k},s_{\vec{u}}}\; .\label{EQ:IDegeneracy}
\end{equation}

%We note that the additional 2-fold degeneracy predicted
%by Eq.~(\ref{EQ:IDegeneracy}) is a consequence\cite{Messiah} of two factors: (i) the Hamiltonian is invariant with
%respect to both space inversion, $[\hat{H}_0,\hat{I}]=0$, and space translation,
%$[\hat{H}_0,\hat{T}_{\vec{R}_n}]=0$; (ii) the fact that the operators $\hat{I}$ and $\hat{T}_{\vec{R}_n}$ do not commute but
%satisfy\cite{Tung} the relation $\hat{I}\hat{T}_{\vec{R}_n}\hat{I}^{-1}=\hat{T}_{-\vec{R}_n}$. If these were not fulfilled the space
%inversion invariance may not have lead to twofold degeneracy. For example in atomic systems where the translational invariance is broken, (i) the space inversion invariance only requires the eigenstates to be of definite parity but does not imlpy additional two-fold degeneracy and (ii) the consequences of time-reversal invariance are themselves equivalent to overall rotational degeneracy.

Expression (\ref{EQ:H0ITransformed})
shows that the states $|\vec{k},s_{\vec{u}}\rangle$ and $|-\vec{k},s_{\vec{u}}\rangle$, respectively belonging to the subspaces
$\varepsilon_{\theta}$ and $\varepsilon_{-\theta}$, must belong to the same degenerate eigenvalue $E^0_{k}$ as a
consequence of the space inversion invariance of $\hat{H}_0$. Taking into account the consequences of the time-reversal invariance
of $\hat{H}_0$ given in Eqs.~(\ref{EQ:HOKTransformed}) and (\ref{EQ:KDegeneracy}), all eigenenergies $E^0_{k}$ of $\hat{H}_0$
must be 4-fold degenerate. To every energy value $E^0_{k}$ corresponds a four-dimensional subspace $\varepsilon^0_{k}$,
which is a direct sum, $\varepsilon^0_{k}=\varepsilon^0_{\theta}+\varepsilon^0_{-\theta}$, of the two subspaces
$\varepsilon^0_{\theta}$ and $\varepsilon^0_{-\theta}$. The subspace $\varepsilon^0_{k}$ is invariant and reducible
with respect to space inversion and time-reversal written symbolically as
$\hat{K}\varepsilon^0_{k}=\varepsilon^0_{k}$ and $\hat{I}\varepsilon^0_{k}=\varepsilon^0_{k}$.

Combination of the space inversion and time-reversal invariance of the Hamiltonian $\hat{H}_0$ of the considered
 \emph{translational invariant} system is equivalent to spin degeneracy.
%\footnote{Note that this is true only for translational invariant systems. In atomic systems where the translational invariance is broken (i) the space inversion invariance only requires the eigenstates to be of definite parity but does not imply additional two-fold degeneracy and (ii) the consequences of time-reversal invariance are itself equivalent to overall rotational degeneracy.}
Formally this is illustrated using the conjugation operator $\hat{C}=\hat{K}\hat{I}$ which commutes
with the Hamiltonian $\hat{H}_0$, $[\hat{H}_0,\hat{C}]=0$ if it commutes separately with $\hat{K}$ and $\hat{I}$. Using the usual procedure of
applying the operator $\hat{C}$ to Eq.(\ref{EQ:H0Eigen}) and taking into account Eq.(\ref{EQ:CPsi}) one obtains the
identity $\hat{H}_0|\vec{k},-s_{\vec{u}}\rangle=E^0_{\vec{k},s_{\vec{u}}}|\vec{k},-s_{\vec{u}}\rangle$. Therefore the Bloch states $|\vec{k},s_{\vec{u}}\rangle$ and $|\vec{k},-s_{\vec{u}}\rangle$
describing an electron with opposite spins belong to the same degenerate energy value $E^0_{k}$. The subspace $\varepsilon^0_{k}$
corresponding to $E^0_{k}$ is invariant with respect to $\hat{C}$. The spin degeneracy can be viewed also as a consequence
of the $SU(2)$ invariance of the Hamiltonian $\hat{H}_0$.

\subsection{Broken Spin Degeneracy}\label{APP:BrokenSpinDegeneracy}
%Note that the requirement for \emph{discrete translational} invariance is significant since broken space inversion invariance is not compatible with infinitesimal translational invariance.
The problem for the spectrum of the Hamiltonian $\hat{H}$ shown in Eq~(\ref{EQ:HamBrokenInversion}) is given by the identity
\begin{equation}
\hat{H}|\vec{\kappa},\sigma_{\vec{u}}\rangle\equiv E_{\vec{\kappa},\sigma_{\vec{u}}}|\vec{\kappa},\sigma_{\vec{u}}\rangle\; .\label{EQ:HEigen}
\end{equation}
The Bloch states $|\vec{\kappa},\sigma_{\vec{u}}\rangle$ are the common set of eigenstates of the commuting operators $\hat{H}$ and $\hat{T}_{\vec{R}_n}$ and $E_{\vec{\kappa},\sigma_{\vec{u}}}$ are the corresponding
eigenvalues of the Hamiltonian $\hat{H}$. They transform among each other according to relations (\ref{EQ:KPsi}), (\ref{IPsi}) and (\ref{EQ:CPsi}) because the Hamiltonian $\hat{H}$ given in Eq~(\ref{EQ:HamBrokenInversion}) is translational invariant.

The spectrum (\ref{EQ:HEigen}) of the Hamiltonian $\hat{H}$ possesses the time-reversal induced properties derived from Eq.~(\ref{EQ:HOKTransformed}) and Eq.~(\ref{EQ:KDegeneracy}).
It is Kramers degenerate and each of its eigenenergies are two-fold degenerate, $E_{\vec{\kappa},\sigma_{\vec{u}}}=E_{-\vec{\kappa},-\sigma_{\vec{u}}}$. However because the space inversion invariance of
the electrostatic potential $\phi(r)$, and therefore of $\hat{H}$
is broken, the degeneracy
due to space inversion invariance is lifted, $E_{\vec{\kappa},\sigma_{\vec{u}}} \neq E_{-\vec{\kappa},\sigma_{\vec{u}}}$.  This requires the lifting of spin degeneracy
\begin{equation}
E_{\vec{\kappa},\sigma_{\vec{u}}}\neq E_{\vec{\kappa},-\sigma_{\vec{u}}}\ .\label{EQ:BrokenSpinDegen}
\end{equation}
The detailed proof follows in the next paragraph.

The broken spin degeneracy is proved by applying the conjugation operator $\hat{C}$ to Eq.~(\ref{EQ:HEigen}). However, because $[\hat{H},\hat{I}]\neq0$
and hence $[\hat{H},\hat{C}] \neq 0$ we can not interchange the positions of $\hat{C}$ and $\hat{H}$. Instead, using
$\hat{1}=C^{-1}C$ and $|\vec{\kappa},-\sigma_{\vec{u}}\rangle=\hat{C}|\vec{\kappa},\sigma_{\vec{u}}\rangle$ one obtains from Eq.~(\ref{EQ:CPsi}) that
\begin{equation}
\hat{C}\hat{H}\hat{C}^{-1}|\vec{\kappa},-\sigma_{\vec{u}}\rangle=E_{\vec{\kappa},\sigma_{\vec{u}}}|\vec{\kappa},-\sigma_{\vec{u}}\rangle\; .\label{EQ:CHIbroken}
\end{equation}
The Bloch state $|\vec{\kappa},-\sigma_{\vec{u}}\rangle$, as an eigenfunction of $\hat{T}_{\vec{R}_n}$, and because $[\hat{H},\hat{T}_{\vec{R}_n}]=0$,
is still an eigenstate of $\hat{H}$ with an eigenvalue $E_{\vec{\kappa},-\sigma_{\vec{u}}}$. However, because of broken inversion invariance,
$$
\hat{C}\hat{H}\hat{C}^{-1}=\hat{I}\hat{K}\hat{H}\hat{K}^{-1}\hat{I}^{-1}=\hat{I}\hat{H}\hat{I}^{-1}\neq\hat{H}$$
$|\vec{\kappa},-\sigma_{\vec{u}}\rangle$ is not anymore an eigenstate of $\hat{H}$ with the eigenvalue $E_{\vec{\kappa},\sigma_{\vec{u}}}$. Instead the Bloch state $|\vec{\kappa},-\sigma_{\vec{u}}\rangle$
is eigenstate of some other operator $\hat{H}'=\hat{I}\hat{H}\hat{I}^{-1}$ with the eigenvalue $E_{\vec{\kappa},\sigma_{\vec{u}}}$. Therefore the states $|\vec{\kappa},\sigma_{\vec{u}}\rangle$
and $|\vec{\kappa},-\sigma_{\vec{u}}\rangle$ do not correspond to the same energy. Electrons in Bloch states characterized by the same
wavevector $\vec{\kappa}$ but having opposite spin orientations do not posses the same energy, the result given in Eq.~(\ref{EQ:BrokenSpinDegen}). %Alternative proof using \emph{reductio ad absurdum} can easily be provided\footnote{T. G. Tenev, PhD Dissertation "Modeling of Electroluminescence in InSb Quantum Wells and Inversion Asymmetric Effects", Lancaster University, 2011}.

\subsection{Discussion of Symmetry Argument}\label{SUBSEC:SymmetryDiscussion}
In the well known treatment\cite{Kittel,DresselhausBand,Zawadski,Winkler} it is supposed that the spin-orbit coupling is part of the model Hamiltonian and that the spin degeneracy
is lifted by the spin-orbit coupling term
\begin{equation}
H_{SO}=\frac{\hbar}{4m^2c^2}\hat{\vec{\sigma}}\cdot\left(\nabla V(\vec{r})\times\hat{\vec{p}}\right)\label{EQ:SO}
\end{equation}
if the electrostatic potential
$V(\vec{r})=V_0(\vec{r})-e\phi(\vec{r})$ does not possess space inversion invariance. However, close examination of the symmetry analysis developed in the text above shows that there is no such requirement. The symmetry analysis is valid also for a nonrelativistic model that contains just electrostatic fields $\phi_0(\vec{r})$ and $\phi(\vec{r})$ and \emph{does not contain
spin-orbit coupling} or magnetic field. It suggests that electrostatic field alone without the mediation of spin-orbit coupling can lead to lift of spin degeneracy given
that $\phi_0(\vec{r})$ and $\phi(\vec{r})$ possess discrete translational invariance and are characterized by preserved and broken space inversion symmetry respectively.

\section{Novel argument for introduction of spin-orbit coupling term}\label{SEC:NewArgument}
Taking spin into account suggests examining for SU(2) symmetry. Since we consider models of a nonrelativistic electron in electrostatic field neglecting spin-orbit coupling,
the Hamiltonians $\hat{H}_0$ and $\hat{H}$ commute with every component of the spin vector operator $\hat{\vec{S}}$ and therefore the Hamiltonians $\hat{H}_0$ and $\hat{H}$
are SU(2) invariant. They commute, $[\hat{R}^s_{\vec{u}}(\phi),\hat{H}]=0$, with every spin rotation operator\cite{Messiah} $\hat{R}^s_{\vec{u}}(\phi)=e^{-i\phi\hat{S}_{\vec{u}}}$
for arbitrary axis $\vec{u}$ and angle of rotation $\phi$. As a consequence of the SU(2) invariance, the spin degeneracy for the model with Hamiltonians $\hat{H}_0$ and $\hat{H}$
must be preserved, in particular for $\hat{H}$
\begin{equation}
E_{\vec{\kappa},\sigma_{\vec{u}}}= E_{\vec{\kappa},-\sigma_{\vec{u}}}
\end{equation}
This result, however, contradicts the result (\ref{EQ:BrokenSpinDegen})
stemming from symmetry analysis based solely on time-reversal and broken space inversion invariance when spin-orbit coupling is neglected.

A possible way to resolve
this inconsistency is the introduction of a term in the model Hamiltonian that breaks the $SU(2)$ invariance and at the same time is consistent with the different cases of
symmetry analysis involving just time-reversal and space-inversion operators. The spin-orbit coupling term (\ref{EQ:SO}),
which does not commute with any spin-rotation operator\cite{Messiah} $\hat{R}^s_{\vec{u}}(\phi)$, satisfies the above requirements and resolves the noted inconsistency. We interpret this as a novel argument for introduction of the spin-orbit coupling term different
from the usual\cite{Messiah,AdvancedSakurai,Thaller} purely relativistic considerations, which give an incorrect numerical factor by $1/2$. This difference is accounted for by Thomas precession or by taking the nonrelativistic limit of the Dirac equation\cite{Messiah,AdvancedSakurai,Thaller}. Thus a realistic model of electron dynamics requires taking
into account the spin-orbit coupling term as a minimum; otherwise we would encounter the above mentioned inconsistencies. Therefore in all subsequent models treated within perturbation method the spin-orbit coupling term is part of the considered Hamiltonian. As a consequence SU(2) symmetry is always broken and there is no requirement for preservation of spin-degeneracy.

\section{Perturbation Method Treatment}\label{SEC:PerturbationMethod}
A model with spin-orbit coupling does not exclude the possibility that electrostatic field by itself leads to lifting of spin degeneracy suggested by the symmetry
analysis based on time-reversal and space-inversion invariance in Sec.~\ref{SUBSEC:SymmetryDiscussion}. It merely shows that if such splitting exists it will lead to additional numerical factor in the spin splitting
already caused by spin-orbit coupling when the space-inversion invariance of the electrostatic field is broken. Since the crystal wavevector $\vec{k}$ varies in discrete steps we investigate this option using standard stationary
perturbation method for degenerate levels.

The unperturbed Hamiltonian is $\hat{H}_0=\frac{\hat{\vec{p}}^2}{2m}+V_0(\vec{r})$, where $V_0(\vec{r})$ possesses space
inversion invariance, $[V_0,\hat{I}]=0$, while the perturbation $\delta\hat{V}$ consists of an electrostatic potential $\phi(r)$ which is odd with respect to space inversion, $\{\phi(r),\hat{I}\}=0$,
and the spin-orbit coupling terms
\begin{subequations}
\begin{eqnarray}
\hat{U}_1 &=& \frac{\hbar}{4m^2c^2}\hat{\vec{\sigma}}\cdot\left(\nabla V_0(r)\times\hat{\vec{p}}\right)\; , \\
\hat{U}_2 &=& -\frac{e\hbar}{4m^2c^2}\hat{\vec{\sigma}}\cdot\left(\nabla\phi(r)\times\hat{\vec{p}}\right)\; .
\end{eqnarray}
\end{subequations}
Unlike in the standard $\vec{k}\cdot\hat{\vec{p}}$ method where the $\vec{k}=0$ stationary states are
used as unperturbed basis, we use the stationary states  $|\vec{k},s_{\vec{u}}\rangle$ of $\hat{H}_0$ for arbitrary $\vec{k}\neq0$. This choice is naturally suggested by the symmetry analysis above since for $\vec{k}=0$ we have just two-fold Kramers degeneracy that is not lifted as far as time-reversal invariance is preserved. We consider 3D
model of triclinic pedial and triclinic pinacoidal systems in which cases the degeneracy of every energy level $E^0_{k}$ of the unperturbed Hamiltonian $\hat{H}_0$
is exactly four-fold. The corresponding subspace $\varepsilon^0_{k}$ is spanned by the four Bloch states $|\vec{k},s_u\rangle$, $|\vec{k},-s_u\rangle$, $|-\vec{k},s_u\rangle$
and $|-\vec{k},-s_u\rangle$.

The first-order correction $E_{k}^{(1)}$ to the energy eigenvalue $E_{k}^{(0)}$ and the zeroth-order states $|0\rangle$ are determined from the eigenvalue equation
\begin{equation}
\hat{P}_{k}^0\delta\hat{V}\hat{P}_{k}^0|0_{k}\rangle=E^{(1)}|0_{k}\rangle\; , \label{EQ:ZeroOrder}
\end{equation}
where $\hat{P}_{k}^0$ is the projector to the subspace $\varepsilon^0_{k}$ and $\delta\hat{V}=-e\phi(\vec{r})+\hat{U}_1+\hat{U}_2$.% in the chosen basis is explicitly given by
%\begin{eqnarray}
%\hat{P}_{k}^0 & = &|\vec{k},s_{\vec{u}}\rangle\langle s_{\vec{u}},\vec{k}|+|\vec{k},-s_{\vec{u}}\rangle\langle -s_{\vec{u}},\vec{k}| + \nonumber \\
% & + & |-\vec{k},s_{\vec{u}}\rangle\langle s_{\vec{u}},-\vec{k}| + |-\vec{k},-s_{\vec{u}}\rangle\langle-s_{\vec{u}},-\vec{k}|\; .
%\end{eqnarray}
The first-order correction  $E_{k}^{(1)}$ to the energy is determined by the solution of the secular equation
\mbox{$\mathrm{det}\left[\hat{P}_{k}^0\delta V\hat{P}_{k}^0-E_{k}^{(1)}\right]=0$} corresponding to the eigenvalue problem (\ref{EQ:ZeroOrder}).
The condition for time-reversal invariance simplifies the secular equation by introducing
relationships between its matrix elements,
\begin{equation}
\left|\begin{array}{c c c c}
a - E^{(1)}_{k} & c & d & 0 \\
c^* & b - E^{(1)}_{k}  & 0 & d \\
d^* & 0 & b - E^{(1)}_{k} & -c \\
0 & d^* & -c^* & a - E^{(1)}_{k}
\end{array}\right| = 0 \; , \label{EQ:Matrix}
\end{equation}
where the matrix elements $a=a_1+a_2+\alpha$, $b=b_1+b_2+\alpha$, $c=c_1+c_2$ and $d=d_1+d_2+\beta$ have contributions from the three different
perturbing terms. The matrix elements due to the spin-orbit coupling term $\hat{U}_1$ are denoted as $a_1$, $b_1$, $c_1$, $d_1$, the ones due to $\hat{U}_2$
as $a_2$, $b_2$, $c_2$, $d_2$, and the ones due to the electrostatic field $\phi(\vec{r})$ as $\alpha$ and $\beta$,
\begin{subequations}
\label{EQS:MatrixElements}
\begin{eqnarray}
a_1 & = \langle s_{\vec{u}},\vec{k}|\hat{U}_1|\vec{k},s_{\vec{u}}\rangle\;  , & a_2 = \langle s_{\vec{u}},\vec{k}|\hat{U}_2|\vec{k},s_{\vec{u}}\rangle\; , \\
b_1 & = \langle -s_{\vec{u}},\vec{k}|\hat{U}_1|\vec{k},-s_{\vec{u}}\rangle\;  , & b_2 = \langle -s_{\vec{u}},\vec{k}|\hat{U}_2|\vec{k},-s_{\vec{u}}\rangle\; , \\
c_1 & = \langle s_{\vec{u}},\vec{k}|\hat{U}_1|\vec{k},-s_{\vec{u}}\rangle\; , & c_2 = \langle s_{\vec{u}},\vec{k}|\hat{U}_2|\vec{k},-s_{\vec{u}}\rangle \; , \\
d_1 & = \langle s_{\vec{u}},\vec{k}|\hat{U}_1|-\vec{k},s_{\vec{u}}\rangle\; , &
d_2  = \langle s_{\vec{u}},\vec{k}|\hat{U}_2|-\vec{k},s_{\vec{u}}\rangle\; , \\
\alpha & = \langle\vec{k}|\delta V(\vec{r}) |\vec{k}\rangle\; , & \beta = \langle\vec{k}|\delta V(\vec{r})|-\vec{k}\rangle \; .\label{EQ:Beta}
\end{eqnarray}
\end{subequations}
% The secular equation derived
%from Eq.(\ref{EQ:Matrix}) has in general two distinct eigenvalues,
%\begin{equation}
%E^{(1)\pm}_{k}=\frac{1}{2}(a+b)\pm\frac{1}{2}\sqrt{(a-b)^2 + 4|c|^2 + 4|d|^2}\; .
%\end{equation}

Taking into account the properties of the electrostatic potential $\phi(\vec{r})$ and the terms $\hat{U}_1$ and $\hat{U}_2$ leads to further
simplification of the above result. The inversion invariance $\hat{I}\hat{U}_1\hat{I}^+=\hat{U}_1$ requires that $a_1=b_1$, $c_1=0$ and that $d_1$ is a
real number. The condition that the term $\hat{U}_2$ must be odd $\hat{I}\hat{U}_2\hat{I}^+=-\hat{U}_2$ with respect to space inversion invariance, requires $a_2=-b_2$ and
$d_2$ to be purely imaginary, while it does not place any restrictions on $c_2$. The fact that the electrostatic field $\phi(\vec{r})$ is odd with respect to
inversion invariance, $\hat{I}\phi(\vec{r})\hat{I}^+=-\phi(\vec{r})$, requires $\alpha=-\alpha$ and thus $\alpha=0$, while $\beta$ needs
to be purely imaginary. Taking into account these simplifications the first order corrections to the energy take the form
\begin{equation}
E^{(1)\pm}_{k}=a_1\pm\sqrt{a_2^2+|c_2|^2+|d_1+d_2+\beta|^2}\; .\label{EQ:FirstOrder}
\end{equation}
Eq.~(\ref{EQ:FirstOrder}) shows
that the spin-orbit coupling term $\hat{U}_1$ possessing space inversion invariance leads to a shift of the first-order energy levels by the
amount $a_1=\langle s_{\vec{u}},\vec{k}|\hat{U}_1|\vec{k},s_{\vec{u}}\rangle$.

\subsection{Perturbation with Translational Invariance}\label{SUBSEC:NullResult}
The expression (\ref{EQ:FirstOrder}) shows that the four-fold degeneracy of $E_{k}^0$ can be lifted up to two twofold degenerate levels in first-order perturbation
method. The spin splitting caused by the spin-orbit coupling term $\hat{U}_2$, which lacks inversion invariance, is embodied in the matrix elements $a_2$, $c_2$ and
$b_2$ within this approach. The presence of the matrix element $\beta$ in the energy correction (\ref{EQ:FirstOrder}) leaves open the option that the electrostatic field leads to an additional contribution to the spin splitting. %The symmetry arguments based solely on time-reversal
%invariance and broken space inversion invariance allow that electrostatic field can lift spin degeneracy without the mediation of spin-orbit coupling.
However, since the condition $\hat{I}\phi(\vec{r})\hat{I}^+=-\phi(\vec{r})$ requires that either $\beta$ is purely imaginary or 0, the consideration of time-reversal
invariance and broken space inversion invariance within perturbation method does not offer conclusive evidence that $\beta\neq0$.

We have supposed that the nonperturbed and perturbed electrostatic potentials possess discrete
translational invariance, $\hat{T}_{\vec{R}_n}\phi(\vec{r})\hat{T}_{\vec{R}_n}^+=\phi(\vec{r})$.
Consequently, the terms $\hat{U}_1$ and $\hat{U}_2$ also possess discrete translational
invariance, $\hat{T}_{\vec{R}_n}\hat{U}_1\hat{T}_{\vec{R}_n}^+=\hat{U}_1$, and
$\hat{T}_{\vec{R}_n}\hat{U}_2\hat{T}_{\vec{R}_n}^+=\hat{U}_2$ respectively. Using this and the Bloch theorem for
\mbox{$d_2=\langle s_{\vec{u}},\vec{k}|\hat{U}_2|-\vec{k},s_{\vec{u}}\rangle$} we find
\begin{eqnarray}
\langle s_{\vec{u}},\vec{k}|\hat{U}_2|-\vec{k},s_{\vec{u}}\rangle
 = \langle s_{\vec{u}},\vec{k}|\hat{T}_{\vec{R}_n}^+
\hat{U}_2\hat{T}_{\vec{R}_n}|-\vec{k},s_{\vec{u}}\rangle = \nonumber \\ =
e^{2i\vec{k}\cdot\vec{R}}\langle s_{\vec{u}},\vec{k}|\hat{U}_2|-\vec{k},s_{\vec{u}}\rangle,
\end{eqnarray}
which for arbitrary $\vec{k}$ and translation vector $\vec{R}$ can be satisfied only if $d_2=0$. Similar
reasoning shows that $d_1=0$ and $\beta=0$. Thus the first-order energy correction (\ref{EQ:FirstOrder}) takes the form
\begin{equation}
E_{k}^{(1)\pm} = a_1 \pm \sqrt{a_2^2 + |c_2|^2}\; .\label{EQ:FirstOrderFinal}
\end{equation}
Therefore it is not possible to have lifting of spin degeneracy of Bloch states purely by translationally invariant electrostatic
field, which is odd with respect to space inversion invariance, \emph{but the reason for this is the discrete translational
invariance of the perturbing electrostatic field $\phi(\vec{r})$}. This is interesting since in the usual symmetry argument translational invariance is not explicitly considered, nor any attention is devoted to it. %Similarly, the spin-orbit coupling term
%$\hat{U}_1$ possessing space inversion invariance does not give partial contribution to spin splitting because
%the term $d_1=0$ due to the translational invariance of the underlying electrostatic potential.
Furthermore, all energy corrections of higher order disappear due to translational invariance since all
$\langle s_{\vec{u}},\vec{k}|\hat{U}|\vec{k}',s_{\vec{u}}\rangle=0$. This makes Eq.~(\ref{EQ:FirstOrderFinal})
exact to all orders in the perturbation method treatment.

% However if the perturbing electrostatic field does not posses translational invariance on top of the % broken space inversion invariance then the matrix element stemming from it are not necessary zero.  % Such electrostatic fields in a crystal can be created by applying external electrostatic field or
% due to built-in potential across p-n junctions. Lets designate such perturbing electrostatic field
% as $\phi'(\vec{r})$ and the matrix element not vanishing due to broken space inversion invariance as % $\beta'=\langles_{\vec{u}},\vec{k}|V'|-\vec{k},s_{\vec{u}}\rangle$.
\subsection{Perturbation without Translational Invariance}
Natural continuation of the above line of thought is to
consider perturbing electrostatic field $\phi'(\vec{r})$ with corresponding energy $V'(\vec{r})=-e\phi'(\vec{r})$ not possessing a center of inversion, $\hat{I}V'(\vec{r})\hat{I}^+\neq V'(\vec{r})$,
and \emph{lacking translational invariance}
\begin{equation}
\hat{T}_{\vec{R}_n}V'(\vec{r})\hat{T}_{\vec{R}_n}^+\neq V'(\vec{r})\; , \label{EQ:BrokenTran}
\end{equation}
to which we will refer as external. Its corresponding
first-order matrix element which is not necessarily zero due to time-reversal invariance is $\beta'=\langle s_{\vec{u}},\vec{k}|\hat{V}'|-\vec{k},s_{\vec{u}}\rangle=\langle -s_{\vec{u}},\vec{k}|\hat{V}'|-\vec{k},-s_{\vec{u}}\rangle$. It is defined between zeroth-order states with spin, but since the electrostatic field does
not contain operator acting directly on the spin-degree of freedom it can be written shorthanded as $\beta'=\langle \vec{k}|\hat{V}'|-\vec{k}\rangle$.
Because $V'(\vec{r})$ is not invariant with respect to space inversion,
the disappearance of the matrix element $\beta'$ is not guaranteed. Examples of electrostatic fields possessing these properties are the
built-in potential in p-n junctions and externally applied electric fields. These are usually orders of magnitude smaller than
the bulk electrostatic fields.

By using the Bloch theorem and the inequality (\ref{EQ:BrokenTran}) we
obtain the inequality
\begin{equation}
\langle s_{\vec{u}},\vec{k}|V'(\vec{r})|-\vec{k},s_{\vec{u}}\rangle\neq e^{2i\vec{k}\cdot\vec{R}}\langle s_{\vec{u}},\vec{k}|V'(\vec{r})|-\vec{k},s_{\vec{u}}\rangle\; ,
\end{equation}
for the matrix element $\beta'$. Therefore for any given $\vec{k}$ and arbitrary translation vector $\vec{R}$ we must have \mbox{$\beta'\neq e^{2i\vec{k}\cdot\vec{R}}\beta'$}.
This condition excludes the possibility the matrix element to be equal to zero and thus $\beta'\neq0$.

In calculating the first-order corrections to the energy we neglect the matrix elements $a'$, $b'$ and $c'$ stemming from the spin-orbit coupling term
\mbox{$\frac{\hbar}{4m^2c^2}\hat{\vec{\sigma}}\cdot\left(\nabla V'(\vec{r})\times\hat{\vec{p}}\right)$} since they should be much smaller than
\mbox{$\beta'=\langle s_{\vec{u}},\vec{k}|\hat{V}'|-\vec{k},s_{\vec{u}}\rangle$} because of the prefactor $\frac{\hbar}{4m^2c^2}$. Since $\beta'\neq0$ it will appear in the expressions for the energy splitting. For the triclinic pedial system in which the bulk potential does not
possess space inversion invariance, the first-order corrections to the energy, \mbox{$E_{k}^{(1)\pm}=a_1\pm\sqrt{a_2^2+|c_2|^2+|\beta'|^2}$}, contain contributions from
the spin-orbit coupling $a_2$ and $c_2$ and a contribution purely from electrostatic field in $\beta'$. The
original fourfold degeneracy is lifted to two distinct two-fold degenerate levels. The energy splitting between them is given by
\begin{equation}
\Delta E_{k}^{(1)} = 2\sqrt{a_2^2 + |c_2|^2 + |\beta'|^2}\; .\label{EQ:SpinSplittingA}
\end{equation}

For the triclinic pinacoidal system where the bulk electrostatic potential possesses space inversion invariance, $a_2$ and $c_2$ are
identically zero. The expression for the first-order energy corrections takes the form $E_{k}^{(1)\pm}=a_1\pm|\beta'|$. The energy splitting,
\begin{equation}
\Delta E_{k}^{(1)}=2|\beta'| \; ,\label{EQ:SpinSplittingB} %=2\langle\vec{k}|V'(\vec{r})|-\vec{k}\rangle
\end{equation}
between the
two twofold degenerate levels has contributions only directly from the electrostatic field through $\beta'$.

\subsection{The lifted degeneracy}
A heuristic argument supporting the interpretation of the energy splittings (\ref{EQ:SpinSplittingA}) and (\ref{EQ:SpinSplittingB}) as spin splitting is
the usual interpretation of the energy splitting due to spin-orbit coupling in systems lacking space inversion invariance\cite{DresselhausBand,Zawadski,Winkler,Kittel} as
spin splitting. This is embodied in the matrix elements $a_2$ and $c_2$ in Eq.~(\ref{EQ:SpinSplittingA}). The additional effect
due to electrostatic field with broken space inversion and translational invariance is embodied in $\beta'$ in Eq.~(\ref{EQ:SpinSplittingA}). Of course because of broken translational invariance of $V'(\vec{r})$ the perturbed states $|\Psi\rangle$ are no longer Bloch states.

The energy splittings (\ref{EQ:SpinSplittingA}) and (\ref{EQ:SpinSplittingB}) can be interpreted as pure spin-splitting for the zero-th order Bloch states. This is easily demonstrated using the invariances of the subspaces $\varepsilon_{k\pm}^{(0)}$ as shown in the following paragraphs.

Since the fourfold degeneracy of $E_{k}^{(0)}$ is not completely removed, each of the first-order corrections
to energy $E_{k\pm}^{(1)}$ corresponds\cite{Messiah} to a two-dimensional subspace $\varepsilon_{k\pm}^{(0)}$ of the
four-dimensional subspace $\varepsilon_{k}^{(0)}$ of the unperturbed problem (\ref{EQ:H0Eigen}). The subspaces
$\varepsilon_{k\pm}^{(0)}$ are mutually exclusive and their direct sum $\varepsilon_{k+}^{(0)}+\varepsilon_{k-}^{(0)}=
\varepsilon_{k}^{(0)}$ is the four-dimensional subspace $\varepsilon_{k}^{(0)}$. The zeroth-order state
$|0_{k}\rangle$, which is the projection of the perturbed state $|\Psi\rangle$ onto the subspace
$\varepsilon_{k}^{(0)}$, cannot be determined uniquely, only its belonging to one of the subspaces
$\varepsilon_{k\pm}^{(0)}$ can be inferred from Eq.~(\ref{EQ:ZeroOrder}).

Symmetry arguments suggest that Kramers degeneracy is preserved and therefore the remaining
twofold degeneracy in first order is the Kramers degeneracy. In order to show this we suppose the equation for first-order energy correction to be solved and consider it as identity
\begin{equation}
\hat{P}_{k}^{(0)}V'(\vec{r})\hat{P}_{k}^{(0)}|0^{\pm}\rangle\equiv E_{k\pm}^{(1)}|0^{\pm}\rangle\label{EQ:ZeroOrderIdentity}\; ,
\end{equation}
where the states $|0^{\pm}\rangle$ are arbitrarily chosen zeroth-order states belonging to the subspaces
$\varepsilon_{k\pm}^{(0)}$, respectively, and therefore to $\varepsilon_{k}^{(0)}$. Applying the time reversal
operator to Eq.(\ref{EQ:ZeroOrderIdentity}) and using that $[\hat{K},\hat{P}_{k}^{(0)}]=0$, because $\varepsilon_{k}^0$
is invariant with respect to $\hat{K}$, $[\hat{K},V'(\vec{r})]=0$ by hypothesis and $E_{k\pm}^{(1)}$ is real, we obtain the identity $\hat{P}_{k}^{(0)}V'(\vec{r})\hat{P}_{k}^{(0)}\hat{K}|0^{\pm}\rangle
\equiv E_{k\pm}^{(1)}\hat{K}|0^{\pm}\rangle$. The $\hat{K}$-transformed state $\hat{K}|0^{\pm}\rangle$ is
orthogonal\cite{Messiah} to $|0^{\pm}\rangle$ since we consider a single-electron system taking into account the spin degree
of freedom . As shown above, $\hat{K}|0^{\pm}\rangle$ also identically satisfies Eq.~(\ref{EQ:ZeroOrderIdentity}) with the eigenvalues
$E_{k\pm}^{(1)}$ and therefore belong to the subspaces $\varepsilon_{k\pm}^{(0)}$, respectively. Since we choose
$|0^{\pm}\rangle$ arbitrarily, and by the above argument every $\hat{K}|0^{\pm}\rangle$ belongs to $\varepsilon_{k\pm}^{(0)}$,
the subspaces  $\varepsilon_{k\pm}^{(0)}$ are invariant
with respect to the time-reversal operator $\hat{K}$, $\hat{K}\varepsilon_{k\pm}^{(0)}=\varepsilon_{k\pm}^{(0)}$.
Thus the remaining two-fold degeneracy in first order is precisely the Kramers degeneracy.

The two-dimensional subspaces $\varepsilon_{k\pm}^{(0)}$ are, however, not invariant with respect to the space-inversion
operator because the perturbation $V'(\vec{r})$ does not commute with it, $[\hat{I},V'(\vec{r})]\neq0$. This is proved
by again applying the time-reversal operator $\hat{I}$ to the identity (\ref{EQ:ZeroOrderIdentity}) satisfied by arbitrary
state $|0^{\pm}\rangle\in\varepsilon_{k\pm}^{(0)}$. However, since $[\hat{I},V'(\vec{r})]\neq0$ and by using,
$\hat{I}^{-1}\hat{I}=1$,  Eq.(\ref{EQ:ZeroOrderIdentity}) transforms into the identity, $\hat{P}_{k}^{(0)}\hat{I}V'(\vec{r})\hat{I}^{-1}\hat{P}_{k}^{(0)}|0^{\pm}\rangle
=E_{k\pm}^{(1)}\hat{I}|0^{\pm}\rangle$ for the $\hat{I}$-transformed state $\hat{I}|0^{\pm}\rangle$. This shows that
$\hat{I}|0^{\pm}\rangle$ does not satisfy Eq.(\ref{EQ:ZeroOrderIdentity}) with eigenvalue $E_{k\pm}^{(1)}$ but a
different equation with the eigenvalue $E_{k\pm}^{(1)}$ because $V'\neq\hat{I}V'\hat{I}^{-1}$.
Therefore the $\hat{I}$-transformed states $\hat{I}|0^{\pm}\rangle$ do not belong to the subspaces $\varepsilon_{k\pm}^{(0)}$.
The choice of $|0^{\pm}\rangle$ is arbitrary apart from the condition $|0^{\pm}\rangle\in\varepsilon_{k\pm}^{(0)}$ and
therefore for \emph{every state} $|0^{\pm}\rangle$ belonging to $\varepsilon_{k\pm}^{(0)}$ the $\hat{I}$-transformed state
$\hat{I}|0^{\pm}\rangle$ \emph{does not belong} to $\varepsilon_{k\pm}^{(0)}$. However ,the four-dimensional subspace
$\varepsilon_{k}^{(0)}=\varepsilon_{k+}^{(0)}+\varepsilon_{k-}^{(0)}$ is invariant with respect to $\hat{I}$, $\hat{I}\varepsilon_{k}^{(0)}=
\varepsilon_{k}^{(0)}$ and therefore $\hat{I}|0^{\pm}\rangle$ must belong to $\varepsilon_{k}^{(0)}$.
Since $\hat{I}|0^{\pm}\rangle$ does not belong to $\varepsilon_{k\pm}^{(0)}$ the only remaining option is that it belongs
to the other two-dimensional subspace $\varepsilon_{k\mp}^{(0)}$. Thus the space inversion operator $\hat{I}$ maps
every state $|0^+\rangle\in\varepsilon_{k+}^{(0)}$ to a state $\hat{I}|0^+\rangle\in\varepsilon_{k-}^{(0)}$ belonging
to the other two-dimensional subspace $\varepsilon_{k-}^{(0)}$ and vice versa, symbolically written as $\hat{I}\varepsilon_{k\pm}^{(0)}=
\varepsilon_{k\mp}^{(0)}$. In other words the zeroth-order states $|0^{\pm}\rangle$ and $\hat{I}|0^{\pm}\rangle$ belong
to the two different first-order energy corrections $E_{k\pm}^{(1)}$ separated from each other by the energy difference
$\Delta E_{k}^{(1)}$.

%Next we prove that it is exactly spin-degeneracy of the zeroth order states that is lifted by \emph{electrostatic potential}
%$\delta\phi(r)$ characterized by discrete translational invariance and broken inversion invariance without the mediation
%of spin-orbit coupling. We do that utilizing the spin-flip operator $\hat{C}=\hat{I}\hat{K}$.

Now consider any arbitrary zeroth-order state
with some spin polarization belonging to the subspace $\varepsilon_{k+}^{(0)}$ which we denote  as $|0^{+}
\rangle$ and apply the spin-flip operator\cite{Kittel} $\hat{C}$ to obtain $\hat{C}|0^{+}\rangle=\hat{I}\hat{K}|0^{+}\rangle=
\hat{I}|0^{+\prime}\rangle$. From the previous results we know that $|0^{+\prime}\rangle=\hat{K}|0^{+}\rangle$
also belongs to the subspace $\varepsilon_{k+}^{(0)}$, while the $\hat{I}$-transformed state $\hat{C}|0^{+}\rangle=
\hat{I}|0^{+\prime}\rangle$ belongs to the other two-dimensional subspace $\varepsilon_{k-}^{(0)}$. Therefore the
spin-flip operator $\hat{C}$, Eq.~(\ref{EQ:CPsi}),
maps any zeroth-order state $|0_{k}\rangle\in\varepsilon_{k\pm}^{(0)}$ to a state belonging to the other two-dimensional
set $|0_{k}\rangle\in\varepsilon_{k\mp}^{(0)}$ similarly to the space inversion operator $\hat{I}$. By definition
a zeroth-order state corresponds to the subspace $\varepsilon_{k\pm}^{(0)}$ if it satisfies Eq.(\ref{EQ:ZeroOrder}) with
the corresponding eigenvalue $E_{k\pm}^{(1)}$. So the zeroth-order states $|0_{k}\rangle\in\varepsilon_{k\pm}^{(0)}$ and
$\hat{C}|0_{k}\rangle\in\varepsilon_{k\mp}^{(0)}$ characterized by identical quantum numbers but describing opposite
spin orientations belong to the two different first-order energy corrections $E_{k\pm}^{(1)}$. This constitutes the proof.

A question might arise as to why should a matrix element $\beta'$ which is diagonal in spin indices be thought of as a spin splitting
even just for the zero-th order states. A perspective to understand the issue is that indeed if the secular problem to be solved reduces
to $2\times2$ matrix then matrix element diagonal in spin indices can not be interpreted as spin splitting. However we consider a $4\times4$
matrix and the matrix element $\beta'$ while diagonal in spin indices is not diagonal in the $4\times4$ matrix. They are on the third diagonal
and they lead to lifting of the original four-fold degeneracy into two two-fold degeneracies. Above we have presented detailed argument for the
interpretation of these as pure spin splitting for the zeroth order Bloch states.

However the perturbed states $|\Psi\rangle$ are no longer Bloch states because the perturbation $V'(\vec{r})$ breaks the space
translation invariance. As a consequence the crystal wavevector $\vec{k}$ is no longer a good quantum number for the perturbed states
and in higher orders in the perturbation method the lifted degeneracy can not be interpreted purely as spin degeneracy. Instead it can
be interpreted as lifting of some sort of orbital-spin degeneracy similar to the fine structure splitting due to spin-orbit coupling in
atomic systems. This is within the perturbation method treatment.

%That is the energy spin-degeneracy
%is lifted purely by electrostatic potential $\delta\phi(r)$ with spin splitting to first order given by
%\mbox{$\Delta E_{k}^{(1)}=\langle s_{\vec{u}}, k|\delta\hat{V}|k,s_{\vec{u}}\rangle$}.

%\section{Discussion}
%Explicit calculations do not confirm the

%\section{Explicit Calculations}

\section{Discussion}\label{SEC:Discussion}
The primary goal of the study has been to explore the hypothesis whether an electrostatic field can lift the spin degeneracy of Bloch states. Integral new contributions of this have been the novel point for introduction of spin-orbit coupling presented in Sec.~\ref{SEC:NewArgument} and the null result of Sec.~\ref{SUBSEC:NullResult}, which have naturally shaped the presentation. The hypothesis presented in mathematical detail in Sec.~\ref{SUBSEC:SymmetryArgument} and discussed in subsection~\ref{SUBSEC:SymmetryDiscussion} has been inspired by a symmetry argument which to the best of our knowledge is due to Kittel\cite{Kittel}. The original argument has been used to predict the lifting of spin degeneracy in the presence of spin-orbit coupling. In subsection~\ref{SUBSEC:SymmetryDiscussion} we noted that it is valid also for a model of electron moving in an electrostatic field in which we take spin degree into account but neglect spin-orbit coupling. Such a model is easily justified on the ground that in nature there is no electron without spin. %Taking into account the spin-degree of freedom of the electron, by using two-component Pauli spinors to represent electron state, naturally leads to introduction of spin-quantum number of type constant of motion. This is so because the Hamiltonian in presence of electrostatic field commutes with every component of the spin-vector operator $\hat{\vec{S}}$.

The trivial consequence of such a model would be double degeneracy of all levels due to the spin-degree of freedom. In such a model, where spin-orbit coupling is neglected, SU(2) symmetry is preserved as noted in Sec.~\ref{SEC:NewArgument}. Thus the double spin degeneracy of such a model can be viewed also as a consequence of SU(2) symmetry. Of course when spin-orbit coupling is taken into account, SU(2) symmetry is broken\cite{Messiah}, and it is not possible to introduce spin quantum number as quantum number of type constant of motion. However we treat spin-orbit coupling terms as perturbations and in the zeroth-order model the spin can be introduced as a separate quantum number.

The original symmetry argument due to Kittel\cite{Kittel} does not use the irreducible representations of the point groups. We considered formulating the symmetry argument using the irreducible representations of the point groups, but for the particular purpose we concluded that working with the symmetry operators themselves is sufficient. This is so, because we consider a conceptual question, therefore we test the hypothesis on the simplest possible systems which posses the required characteristics of the problem. These are 3D crystals of the triclinic crystal system: the triclinic pedial and triclinic pinacoidal crystal classes. These are the least symmetric classes, which have in common only a rotation by $2\pi$ and a time-reversal symmetry. In addition the pedial class is characterized with a broken space inversion symmetry, while the only other symmetry element of the pinacoidal class is the space inversion.

The compatibility relations of the irreducible representations can be used to calculate the effects, including lifting of degeneracy, by electric fields which do not break the translational invariance. However, this approach would not offer an answer to the hypothesis whether an electrostatic field alone would break the spin degeneracy. This is so because based solely on the symmetry argument we cannot say whether the spin splitting is caused by the electrostatic field alone or by the spin-orbit coupling. This is the reason to use the perturbation method. Indeed, when we use the perturbation method it turns out that if the translational invariance of the electrostatic field is preserved no spin-splitting occurs. On the other hand the perturbation method treatment indicates a lifting of the spin degeneracy when both the translation and inversion invariances are broken. However, if the translational invariance is broken we can no longer talk about space symmetry groups and their subgroups - the point groups. This is so because the discussion of the space groups requires the translational invariance of the crystal lattice.

\section{Conclusions}\label{SEC:Conclusions}
We have scrutinized the symmetry argument based on space inversion and time-reversal invariance predicting the appearance of spin
splitting in case of broken space inversion symmetry. A novel argument for the need for introduction of a term breaking SU(2) invariance like
the spin-orbit coupling term has been presented. This argument is different from the usual arguments based on special relativity used for
the introduction of spin-orbit coupling term. We have shown that in systems possessing discrete translational invariance it is not possible
to have spin splitting solely by electrostatic field with broken space inversion symmetry due to the preserved translational invariance
while a spin splitting exists due to the combination of spin-orbit coupling and electrostatic field. The possibility for lifting of spin degeneracy due to electrostatic field without the mediation of spin-orbit coupling has been investigated using perturbation method treatment suggesting its possible existence in systems characterized by both broken space inversion invariance and broken translational invariance, as far as the perturbation method is applicable. There is a possibility that the last result is a quirk of the application of the perturbation method to the case of electrostatic field with broken translational invariance. The problem might be further clarified and the results tested theoretically by treating the same scenario in the Dirac equation. If they are confirmed the possibility needs to be tested experimentally, most suitable for which maybe systems of the triclinic pinacoidal symmetry system.

%\begin{acknowledgments}
%\end{acknowledgments}

%

%\bibliography{thesis,books,SymmetricRashba}

\end{document}